\documentclass[aps,pre,superscriptaddress,noshowpacs,longbibliography]{revtex4-2}
\usepackage{amsmath}
\usepackage{amssymb}
\usepackage{tikz}
\usepackage{graphicx}
\usepackage{dcolumn}
\usepackage{bm}
\usepackage{epsfig}
\usepackage{psfrag}
\def\II{\hbox{$1\hskip -1.2pt\vrule depth 0pt height 1.6ex width 0.7pt\vrule depth 0pt height 0.3pt width 0.12em$}}
\newcommand{\reffig}[1]{\mbox{Fig.~\ref{#1}}}
\newcommand{\refeq}[1]{\mbox{Eq.~(\ref{#1})}}
\newcommand{\refsec}[1]{\mbox{Sec.~\ref{#1}}}
\newcommand{\reftab}[1]{\mbox{Tab.~\ref{#1}}}
\newcommand{\be}{\begin{equation}}
\newcommand{\ee}{\end{equation}}
\newcommand{\bal}{\begin{align}}
\newcommand{\eal}{\end{align}}
\newcommand{\ba}{\begin{eqnarray}}
\newcommand{\ea}{\end{eqnarray}}

\newcommand{\T}{${\mathcal T}\,$}
\newcommand{\Ti}{${\mathcal T}$}
\def\II{\hbox{$1\hskip -1.2pt\vrule depth 0pt height 1.6ex width 0.7pt\vrule depth 0pt height 0.3pt width 0.12em$}}

\begin{document}

\title{Time-reversal Invariance Violation and Quantum Chaos Induced by Magnetization in Ferrite-Loaded Resonators}

\author{Weihua Zhang}
\address{%
Lanzhou Center for Theoretical Physics and the Gansu Provincial Key Laboratory of Theoretical Physics, Lanzhou University, Lanzhou, Gansu 730000, China
}
\address{Center for Theoretical Physics of Complex Systems, Institute for Basic Science (IBS), Daejeon 34126, Korea}
\author{Xiaodong Zhang}
\address{%
Lanzhou Center for Theoretical Physics and the Gansu Provincial Key Laboratory of Theoretical Physics, Lanzhou University, Lanzhou, Gansu 730000, China
}
\author{Barbara Dietz}
\email{bdietzp@gmail.com}
\address{%
Lanzhou Center for Theoretical Physics and the Gansu Provincial Key Laboratory of Theoretical Physics, Lanzhou University, Lanzhou, Gansu 730000, China
}
\address{Center for Theoretical Physics of Complex Systems, Institute for Basic Science (IBS), Daejeon 34126, Korea}

\date{\today}

\begin{abstract}    
We investigate the fluctuation properties in the eigenfrequency spectra of flat cylindrical microwave cavities that are homogeneously filled with magnetized ferrite. These studies are motivated by experiments in which only small pieces of ferrite were embedded in the cavity and magnetized with an external static magnetic field to induce partial time-reversal (\T) invariance violation. We use two different shapes of the cavity, one exhibiting an integrable wave dynamics, the other one a chaotic one. We demonstrate that in the frequency region where only transverse-magnetic modes exist, the magnetization of the ferrites has no effect on the wave dynamics and does not induce \Ti-invariance violation whereas it is fully violated above the cutoff frequency of the first transverse-electric mode. Above all, independently of the shape of the resonator, it induces a chaotic wave dynamics in that frequency range in the sense that for both resonator geometries the spectral properties coincide with those of quantum systems with a chaotic classical dynamics and same invariance properties under application of the generalized \T operator associated with the resonator geometry.   
\end{abstract}
\bigskip
\maketitle

\section{Introduction}
Flat, cylindrical microwave resonators are used since three decades to investigate in the context of Quantum Chaos~\cite{Brody1981,Zimmermann1988,Guhr1989,Guhr1998,Weidenmueller2009,Gomez2011,Frisch2014,Mur2015,Haake2018} the properties of the eigenfrequencies and wave functions of quantum systems with a chaotic dynamics in the classical limit~\cite{Sridhar1991,Stein1992,Graef1992,Deus1995,StoeckmannBuch2000,Dietz2015}. Here, the analogy to quantum billiards of corresponding shape is used, which holds below the frequency of the first transverse-magnetic mode. Namely, in the frequency range, where the electric field strength is parallel to the cylinder axis, the associated Helmholtz equation is identical to the Schr\"odinger equation of the quantum billiard of corresponding shape. The classical counterpart of a quantum billiard consists of a two-dimensional bounded domain, in which a point-like particle moves freely and is reflected specularly on impact with the boundary~\cite{Sinai1970,Bunimovich1979,Berry1981,LesHouches1989}. The dynamics of the billiard depends only on its shape. Therefore, such systems provide a suitable model to investigate signatures of the classical dynamics in properties of the associated quantum system. Experiments have also been performed with three-dimensional microwave resonators~\cite{Weaver1989,Ellegaard1995,Deus1995,Alt1996,Alt1997,Dembowski2002}, where the analogy to the quantum billiard of corresponding shape is lost, because of the vectorial nature of the Helmholtz equations. Their objective was the study of wave-dynamical chaos.

In the above mentioned experiments properties of quantum systems with a chaotic classical counterpart and preserved time-reversal (\T) invariance were investigated. In order to induce \Ti-invariance violation in a quantum system a magnetic field is introduced. Quantum billiards with partially violated \T invariance were modeled experimentally with flat, cylindrical microwave resonators containing one or more pieces of ferrite that were magnetized with an external magnetic field~\cite{So1995,Wu1998,Schanze2001,Dietz2009,Dietz2010,Dietz2019b,Bialous2020,Zhang2022}. Time-reversal invariance is violated through the coupling of the spins in the ferrite, which precess with the Larmor frequency about the external magnetic field, to the magnetic-field component of the resonator modes, which depends on the rotational direction of polarization of the latter. The effect, that leads to \Ti-invariance violation is especially pronounced in the vicinity of the ferromagnetic resonance and its harmonics or at the resonance frequencies of the ferrite piece that lead to trapped modes inside in it~\cite{Dietz2019b}. The motivation of the present study is the understanding of the electrodynamical processes that take place inside a magnetized cylindrical ferrite, especially of its wave-dynamical properties and dependence on its shape. 

In this work we compute with COMSOL multiphysics the eigenfrequencies and electric field distributions of flat, cylindrical metallic resonators, that are homogeneously filled with fully magnetized ferrite material and investigate the fluctuation properties in the eigenfrequency spectra in the realm of Quantum Chaos. We choose two different shapes, one which has the shape of a billiard with integrable dynamics, the other one with chaotic dynamics. The spectral properties are studied in the frequency range where only transverse-electric modes exist and the Helmholtz equation is scalar, and in the region where it is vectorial. In~\refsec{Fe} we review properties of ferrite and the associated wave equations. They reduce to that of the quantum billiard of corresponding shape with no magnetic field in the two-dimensional case. In~\refsec{Num} we present the models that are investigated and in~\refsec{Res} the results for the spectral properties. Interestingly, in the region where also transverse-magnetic modes exist, even for the resonator with the shape corresponding to a dielectric-loaded cavity with integrable wave dynamics, they coincide with those of a quantum system with chaotic classical dynamics. Our findings are discussed in~\refsec{Concl}.    

\section{Review of the Wave Equations for a Metallic Resonator Homogeneously Filled with Magnetized Ferrite\label{Fe}}
A ferrite is a non-conductive ceramic with a ferrimagnetic crystal structure.  Similar to antiferromagnets, it consists of different sublattices whose magnetic moments are opposed and differ in magnitudes. When applying a static external magnetic field, these magnetic moments become aligned. This process can be effectively described as a macroscopic magnetic moment. We consider a flat cylindrical resonator made of ferrite~\cite{Lax1962,Soohoo1960} which is magnetized by a static magnetic field perpendicular to the resonator plane, and enclosed by a perfect electric conductor (PEC). The macroscopic Maxwell equations for an electromagnetic field with harmonic time variation~\cite{Jackson1999},
\be
\vec E(\vec x,t)=\vec E(\vec x)e^{-i\omega t},\, \vec B(\vec x,t)=\vec B(\vec x)e^{-i\omega t}
\ee
propagating through the resonator read
\ba
        &\vec\nabla\times\vec E&=i\omega\vec B\label{WE1} \\
	&\vec\nabla\times (\hat\mu^{-1}\vec B)&=-i \frac{\epsilon_r(\vec x)}{c^2}\omega\vec E\label{WE2} \\
        &\vec\nabla\cdot\vec B&=0\label{WE3}\\
        &\vec\nabla\cdot\vec E&=0\label{WE4},
\ea
where $c=\frac{1}{\sqrt{\epsilon_0\mu_0}}$ is the speed of light in vacuum, $\epsilon_0$ and $\mu_0$ are the permittivity and permeability in vacuum and $\epsilon_r$ is the relative permittivity, which due to the assumed homogeneity of the material does not depend on $\vec x$ in the bulk of the resonator. Furthermore, $\hat\mu_r$ denotes the permeability tensor, which may be expressed in terms of the susceptibility tensor $\hat\chi$ as $\hat\mu_r=\II+\hat\chi$, and results from the magnetization $\vec M(\vec x)=\hat\chi\vec H(\vec x)$ of the ferrite, $\vec B(\vec x)=\mu_0\hat\mu_r\vec H(\vec x)=\mu_0\left(\vec H(\vec x)+\vec M(\vec x)\right)$. 

Combination of the first two equations yields wave equations for $\vec E(\vec x)$ and $\vec B(\vec x)$~\cite{So1995,StoeckmannBuch2000},
\ba
&\vec\nabla\times\left[\hat\mu_r^{-1}\nabla\times\vec E(\vec x)\right]-\epsilon_r(\vec x)k_0^2\vec E(\vec x)&=0,\label{WE}\\
&\vec\nabla\times\left\{\vec\nabla\times\left[\hat\mu_r^{-1}\vec B(\vec x)\right]\right\}-\epsilon_r(\vec x)k_0^2\vec B(\vec x)&=0,\label{WM}
\ea
with $k_0=\frac{\omega}{c}$ denoting the wavenumber in vacuum. The magnetization follows the equation of motion
\be
\frac{d\vec M}{dt}=\vert\gamma\vert\vec M\times\vec H,
\label{EM}
\ee
resulting from the torque exerted by the magnetic field $\vec H(\vec x)$ on $\vec M(\vec x)$. Here, $\gamma=-2.21\cdot 10^5mA^{-1}s^{-1}$ denotes the gyromagnetic ratio of the electron. The magnetic field $\vec H(\vec x)$ is composed of the static magnetic field $\vec H_0=H_0\vec e_z$ applied in the direction of the cylinder axis of the resonator, which is chosen parallel to the $z$ axis, and the magnetic-field component of the electromagnetic field,
\be
\vec H(\vec x)=H_0\vec e_z+\vec h(\vec x)e^{i\omega t}.
\label{H0}
\ee
We assume that the strength $H_0$ is sufficiently large to ensure that the magnetization attains its saturation value $M_s=\chi_0H_0$, with $\chi_0$ denoting the static susceptibility. Similarly, $\vec M(\vec x)$ is given by a superposition of the static magnetization $M_s\vec e_z$ and the magnetization resulting from the electromagnetic field,
\be
\vec M=M_s\vec e_z+\vec me^{i\omega t}.
\label{M0}
\ee
The magnetization $\hat M$ is obtained by inserting Eqs.~(\ref{H0}) and~(\ref{M0}) into~\refeq{EM}. We may assume that the contributions originating from the electromagnetic field in Eqs.~(\ref{H0}) and~(\ref{M0}) are sufficiently small compared to that of the static parts, so that only terms linear in $\vec h$ and $\vec m$ need to be taken into account, yielding $\vec M(\vec x)=\hat\chi\vec H(\vec x)$ with
\be
\hat\chi=\begin{pmatrix}
        \chi &-i\kappa &0\\
        i\kappa &\chi &0\\
        0 &0 &\chi_0
\end{pmatrix},
\ee
and $\chi(\omega)=\frac{\omega_L\omega_M}{\omega_L^2-\omega^2},\, \kappa(\omega)=\frac{\omega\omega_M}{\omega_L^2-\omega^2}$. Here, $\omega_M=\vert\gamma\vert M$ and $\omega_L=\vert\gamma\vert H_0$ denote the precession frequency about the saturation magnetization $\vec M_s$ and the Larmor frequency, with which the magnetization $\vec M$ presesses about $\vec H_0$, respectively. The quantities $\chi(\omega)$ and $\kappa(\omega)$ exhibit a pronounced resonance behavior around the ferromagnetic resonance $\omega=\omega_L$. With these notations the inverse of $\hat\mu_r$ is given by 
\be
\hat\mu_r^{-1}=\left(\II+\hat\chi\right)^{-1}=\begin{pmatrix}
        \frac{1+\chi}{\delta} &\frac{i\kappa}{\delta} &0\\
        -\frac{i\kappa}{\delta} &\frac{1+\chi}{\delta} &0\\
        0 &0 &\frac{1}{1+\chi_0}
\end{pmatrix},\, \delta=(1+\chi)^2-\kappa^2.
\ee
The resonators under consideration have a cylindrical shape with a non-circular cross section and the external magnetic field is constant and perpendicular to the resonator plane. Furthermore, the ferrite material is homogeneous, that is, in the bulk the entries of $\hat\mu_r$ and $\epsilon_r$ are spatially constant and only depend on the angular frequency $\omega$ of the electromagnetic field. This is distinct from the experiments presented in Refs.~\cite{So1995,Wu1998,Schanze2001,Dietz2009,Dietz2010,Dietz2019b,Bialous2020,Zhang2022}, where \T invariance violation was induced by inserting cylindrical ferrites with circular cross section into an evacuated metallic resonator. There, the origin of the \T invariance violation is the coupling of the spins of the ferrite to the magnetic field components of the electromagnetic field excited in the resonator, which depends on their rotational direction. It is strongest in the vicinity of the ferromagnetic resonances and at resonance frequencies of the ferrite, were modes are trapped in it~\cite{Dietz2019b}. In these microwave resonators, $\hat\mu_r$ and $\epsilon_r$ are spatially dependent, since they experience a jump at the surface of the ferrite.   

For a ferrite enclosed by a PEC the boundary conditions are given by
\be
\vec n(\vec x_S)\times\vec E=\vec 0,\, \vec n(\vec x_S)\cdot\vec B=0,\label{BC}
\ee
with $\vec n(\vec x_S)$ denoting the normal to the ferrite surface at $\vec x_S$ pointing away from the resonator. Here, the surface charge density and surface current density may be neglected due to the high resistivity of the ferrite. Defining $\vec E=E_x\vec e_x+E_y\vec e_y+E_z\vec e_z$, this yields at the bottom and top planes of a flat, cylindrical resonator of height $h$, where $-\vec n(x,y,z=0)=\vec n(x,y,z=h)=\vec e_z$, the boundary conditions 
\be
E_x(x,y,z=0)=E_x(x,y,z=h)=E_y(x,y,z=0)=E_y(x,y,z=h)=0.\label{BC1} 
\ee
Along the side wall they read
\be
n_xE_y=n_yE_x,\, E_z=0,\, \vec n_t=(n_x(s),n_y(s),0)\label{BC2}
\ee
with $s$ parametrizing the contour of the resonator in a plane parallel to the $(x,y)$ plane. This condition leads to a coupling of $E_x$ and $E_y$.

Accordingly, we may separate the electromagnetic field into modes propagating in the resonator plane, denoted by an index $t$ and modes perpendicular to it, i.e., in $z$ direction.
\be
\vec E=\vec E_t+E_z\vec e_z,\, \vec B=\vec B_t+B_z\vec e_z,\, \vec\nabla=\vec\nabla_t+\vec e_z\frac{\partial}{\partial z}
\ee
and
\be
\hat\mu_r^{-1}\vec\nabla=\hat m\vec\nabla_t+m_0\vec e_z\frac{\partial}{\partial z},\,
\hat\mu_r^{-1}\vec B=\hat m\vec B_t+m_0B_z\vec e_z
\ee
with
\be
\hat m=\begin{pmatrix}
        \frac{1+\chi}{\delta} &\frac{i\kappa}{\delta}\\
        -\frac{i\kappa}{\delta} &\frac{1+\chi}{\delta}
\end{pmatrix},\, m_0=\frac{1}{1+\chi_0}.
\ee
Furthermore, due to the cylindrical shape we may assume that
\be
\vec E(x,y,z)=\vec E(x,y)e^{-ik_z z},\, \vec B(x,y,z)=\vec B(x,y)e^{-ik_zz}.
\ee
The electromagnetic waves are reflected at the PECs terminating the resonator at the top and bottom, implying that
\be
k_z=q\frac{\pi}{h},\, q=0,1,\dots ,
\label{kz}
\ee
that is, $k^2=k^2_t +k^2_z=k^2_t+q^2\left(\frac{\pi}{h}\right)^2$ for $k\geq q\frac{\pi}{h}$.

The Maxwell equations become
\begin{align}
i\omega\vec B_t&=\left[ik_z\vec E_t+\vec\nabla_tE_z\right]\times \vec e_z,\,
        & i\omega B_z&=\left[\vec\nabla_t\times\vec E_t\right]\cdot\vec e_z\label{E1}\\
	-i\frac{\epsilon_r(\vec x)}{c^2}\omega\vec E_t&=\left[m_0\vec\nabla_t B_z +ik_z\hat m\vec B_t\right]\times\vec e_z,\,
	& -i\frac{\epsilon_r(\vec x)}{c^2}\omega E_z&=\left[\vec\nabla_t\times\left(\hat m\vec B_t\right)\right]\cdot\vec e_z\label{E2}\\
\vec\nabla_t\cdot\vec E_t&=ik_zE_z,\,
        & \vec\nabla_t\cdot\vec B_t&=ik_zB_z.\label{E3}
\end{align}
The in-plane modes can be expressed in terms of the modes perpendicular to the plane. For this we insert the first equation of~\refeq{E1} into the first one of~\refeq{E2} and vice versa yielding
\ba
-i\left[\epsilon_rk_0^2\II-k^2_z\hat m\right]\vec E_t=&&\omega m_0\vec\nabla_tB_z\times\vec e_z-k_z\hat m\vec\nabla_t E_z\\
i\left[\epsilon_rk_0^2\II-k^2_z\hat m\right]\vec B_t=&&k_z m_0\vec\nabla_tB_z+\frac{\epsilon_r(\vec x)}{c^2}\omega\vec\nabla_tE_z\times\vec e_z.
\ea
The wave equation~\refeq{WE} can also be separated into in-plane modes and modes perpendicular to the resonator plane. Namely,
\be
\vec\nabla\times\left[\hat\mu_r^{-1}\vec\nabla\times\vec E(\vec x)\right]=
\hat\mu_r^{-1}\vec\nabla\left(\vec\nabla\cdot\vec E\right)+\left(\hat\mu_r^{-1}\vec\nabla\right){\overleftarrow\nabla}\cdot\vec E-\vec\nabla\cdot\left(\hat\mu_r^{-1}\vec\nabla\right)\vec E
\ee
where the gradient ${\overleftarrow\nabla}$ is applied to the term to its left. According to~\refeq{WE4} the first term on the right hand side vanishes. Inserting this equation into~\refeq{WE} and separating into modes in the resonator plane and perpendicular to it yields
\ba
&\left[\vec\nabla_t\cdot\left(\frac{1+\chi}{\delta}\vec\nabla_t\right)-ik_z\frac{\partial}{\partial z}\left\{\frac{1}{1+\chi_0}\right\}+i\vec\nabla_t\cdot\left(\vec e_z\times\frac{\kappa}{\delta}\vec\nabla_t\right)\right]\vec E\label{Helmholtz}\\
&-\begin{pmatrix}
        \left(\vec\nabla\left\{\frac{1+\chi}{\delta}\right\}\frac{\partial}{\partial x}-\vec\nabla\left\{\frac{i\kappa}{\delta}\right\}\frac{\partial}{\partial y}\right)\cdot\vec E\\
        \left(\vec\nabla\left\{\frac{1+\chi}{\delta}\right\}\frac{\partial}{\partial y}+\vec\nabla\left\{\frac{i\kappa}{\delta}\right\}\frac{\partial}{\partial x}\right)\cdot\vec E\\
-ik_z\vec\nabla\left\{\frac{1}{1+\chi_0}\right\}\cdot\vec E
\end{pmatrix}
=\left(\frac{1}{1+\chi_0}k^2_z-\epsilon_rk_0^2\right)\vec E,
\ea
where curly brackets mean that $\vec\nabla$ is only applied to the terms framed by them. For $q=0$ in~\refeq{kz}, i.e., $k_z=0$ the electric field is perpendicular to the resonator plane, $\vec E(\boldsymbol{r})=E(x,y)\vec e_z$ and~\refeq{Helmholtz} becomes
\be
\left[\vec\nabla_t\cdot\left(\frac{1+\chi}{\delta}\vec\nabla_t\right)+i\vec\nabla_t\cdot\left(\vec e_z\times\frac{\kappa}{\delta}\vec\nabla_t\right)\right]E(x,y)
=-\epsilon_rk_0^2E(x,y),\, E(x,y)\big\vert_{\partial\Omega}=0\label{Helm1}
\ee
with Dirichlet boundary conditions along the boundary $\partial\Omega$. For the case considered here, i.e., for spatially constant $\hat\mu_r$, the wave equation reduces to the scalar Helmholtz equation
\be
\Delta_tE(x,y)=-\epsilon_r\frac{\delta}{1+\chi}k^2_0E(x,y)=-k^2E(x,y),\, E(x,y)\big\vert_{\partial\Omega}=0\label{Helm2}
\ee
yielding the dispersion relation
\be
k=\sqrt{\epsilon_r\frac{\delta}{1+\chi}}k_0=\sqrt{\epsilon_r\frac{(\omega_L+\omega_M)^2-\omega^2}{\omega_L(\omega_L+\omega_M)-\omega^2}}k_0.\label{Dispersion}
\ee
Equations~(\ref{Helm1})-~(\ref{Dispersion}) hold up to
\be
k^{crit}=\frac{\pi}{h}
\ee
or, equivalently, with $\bar\omega^2 = \frac{(\omega_L+\omega_M)^2+\left(\frac{c\pi}{h\sqrt{\epsilon_r}}\right)^2}{2}$
\be
\omega^{crit}=\sqrt{\bar\omega^2\pm\sqrt{\bar\omega^4-\omega_L(\omega_M+\omega_L)\left(\frac{c\pi}{h\sqrt{\epsilon_r}}\right)^2}},
\ee
where for vanishing external field, i.e., for $\omega_L=\omega_M=0$ the plus sign has to be taken in the radicand, yielding
\be
\omega_r^{crit}=\frac{c\pi}{h\sqrt{\epsilon_r}}.
\ee
For nonzero $\omega_L$ and $\omega_M$ the minus sign applies. Thus, below the cutoff circular frequency of the first transverse-electric mode, referred to as critical in the following, $\omega^{crit}$ the wave equation coincides with that of a quantum billiard~\cite{Sinai1970,Bunimovich1979,LesHouches1989} in a dispersive medium~\cite{Lebental2007,Bogomolny2008,Bittner2012a,Bittner2012b}. This correspondence between quantum billiards and the scalar Helmholtz equation of flat, cylindrical microwave cavities has been used in numerous experiments to determine their eigenvalues and eigenfunctions~\cite{Sridhar1991,Graef1992,Stein1992,StoeckmannBuch2000,Dietz2015}. The corresponding classical billiard consists of a point particle which moves freely inside a bounded two-dimensional domain and is reflected specularly at the wall. 

\section{Numerical Analysis\label{Num}}
\begin{figure}[ht!]
\centering
\includegraphics[width=0.3\linewidth]{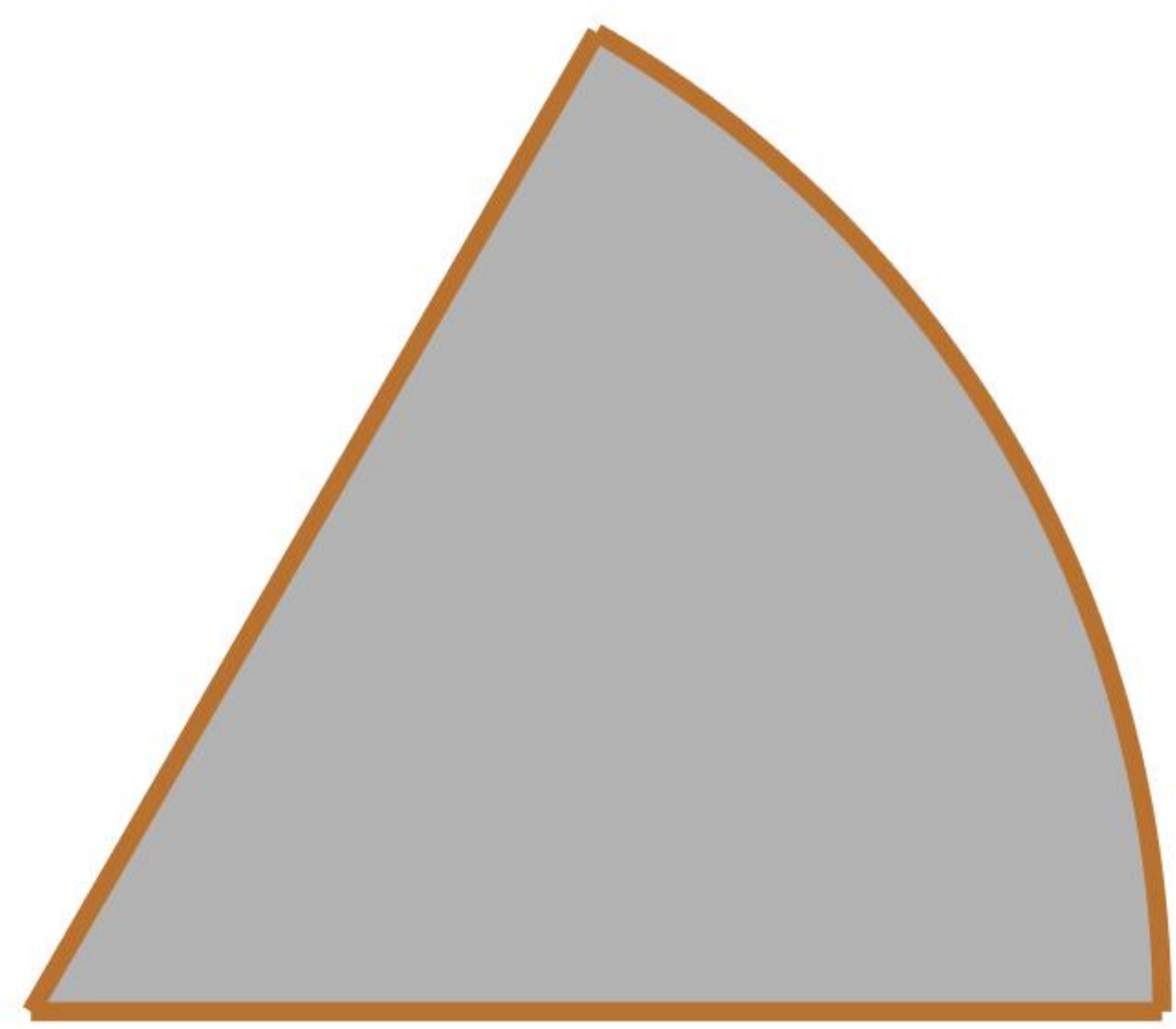}
\includegraphics[width=0.3\linewidth]{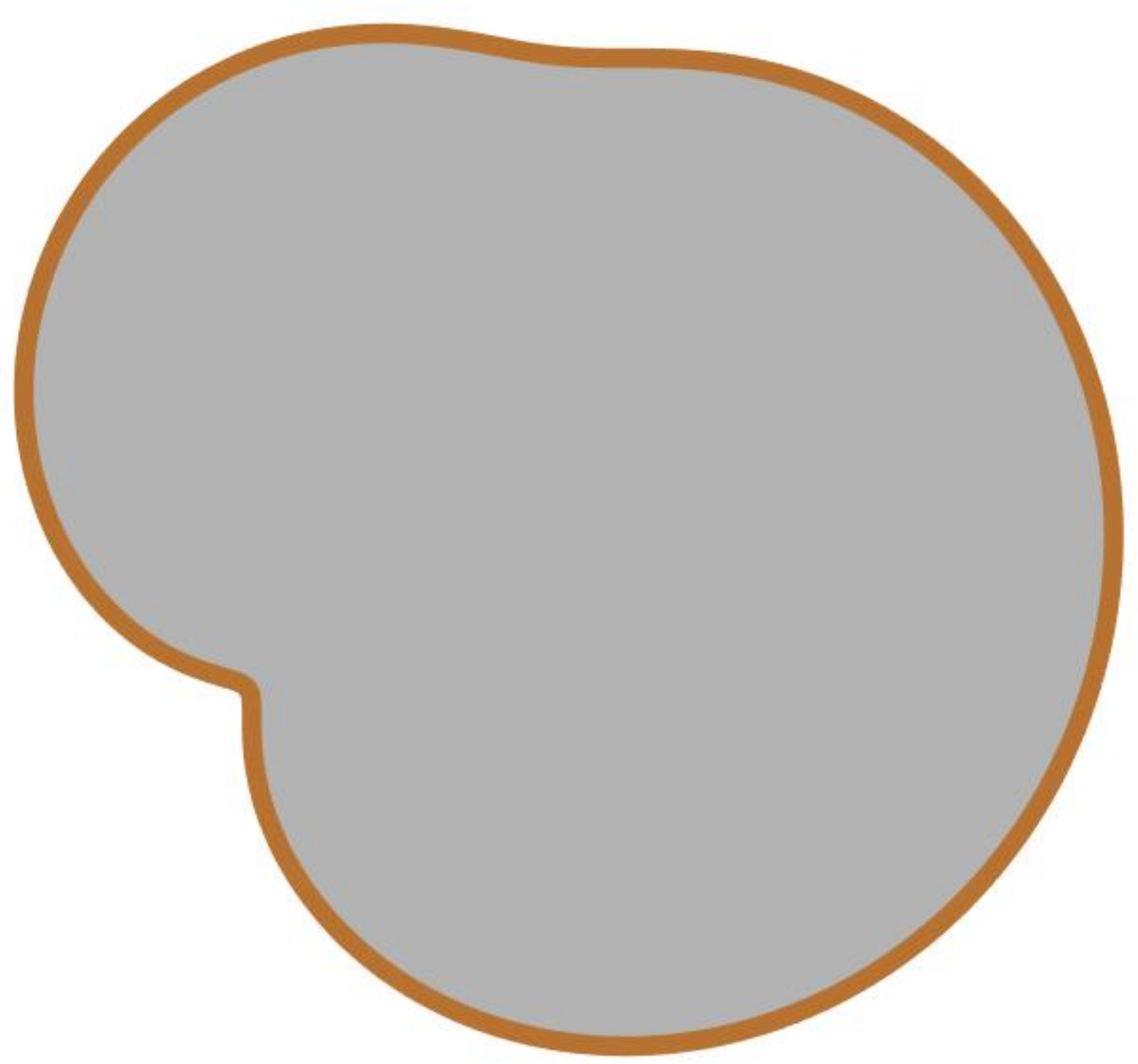}
	\caption{Sketch of the ferrite-loaded cavities, which have the shape of a circle-sector billiard (left) with inner angle $60^\circ$ and of a Africa billiard (right). The cavity is made of a PEC (brown lines), and the filling consists of magnetized ferrite (gray domain).} 
\label{fig1}
\end{figure}

We investigated the spectral properties of ferrite-loaded metallic resonators with the shapes shown in~\reffig{fig1}, using COMSOL multiphysics. We set the properties of the ferrite material to those of $18 \rm{G_3}$ ferrite from the Y-Ga-In series, which has a low loss, of which the relative permittivity and saturation magnetization are $\varepsilon_r=14.5$ and $M_s=1.47\cdot 10^5 \rm{A/m}$, respectively. The other parameters are given in~\reftab{tab1}. 
\begin{table}[h!]
        \begin{tabular}{|c|c|c|c|c|c|c|}
                \hline
		Shape  &h  &Area $\mathcal{A}$    & $\omega_L$ & $\omega_M$ & $\omega_r^{crit}$ &$\omega^{crit}$\\ \hline
		Sector &20~mm & 0.3355~m$^2$  & 87.96~GHz          & 32.49~GHz           & 24.71~GHz  &10.56~GHz            \\ \hline
		Africa &10~mm & 0.0377~m$^2$ & 43.98~GHz          & 32.49~GHz           & 12.35~GHz  &18.34~GHz             \\ \hline
        \end{tabular}
        \caption{Parameters for the two resonator realizations.}
        \label{tab1}
\end{table}

The sector has a radius of 800~mm. The wave dynamics of microwave resonators with this shape is integrable~\cite{Weaver1989,Ellegaard1995,Deus1995,Alt1996,Alt1997}. The boundary of the Africa shape $[x(r,\varphi),y(r,\varphi)]$ is defined in the complex plane $w(r,\varphi)=x(r,\varphi)+iy(r,\varphi)$ by 
\be
w(r,\varphi)=r_0\left(z+0.2z^2+0.2z^3e^{i\pi/3}\right), 
\ee
with $z=re^{i\varphi}$ and $r_0 =100$~mm. 
Below $\omega^{crit}$ the wave equation~\refeq{Helmholtz} reduces to the Schr\"odinger equation for the quantum billiard of corresponding shape; see~\refeq{Helm2}. For the sector quantum billiard the solutions of~\refeq{Helm2}, namely, the eigenvalues $k_{p,\nu}$ and eigenfunctions $\Psi_{p,\nu}(r,\varphi)$ are known,
\be
\Psi_{p,\nu}(r,\varphi)=\sin\left(\frac{3}{2}p\varphi\right)J_{\frac{3}{2}p}(k_{p,\nu}r),\, J_{\frac{3}{2}p}(k_{p,\nu}r_0)=0,
\ee
whereas for the Africa billiard the dynamics is fully chaotic~\cite{Berry1986} so that they need to be computed numerically, e.g., with the boundary integral method~\cite{Baecker2003}. We computed the eigenstates with COMSOL multiphysics which employes a finite element method using the parameters listed in~\reftab{tab1}. Note, that beyond the critical frequency $\omega\gtrsim\omega^{crit}$, where the Helmholtz equation becomes three-dimensional, the analogy to the three-dimensional quantum billiard of corresponding shape is lost. 

\section{Results\label{Res}}
\subsection{Electric-Field Distributions}
\begin{figure}[htbp]
\centering
\includegraphics[width=0.5\linewidth]{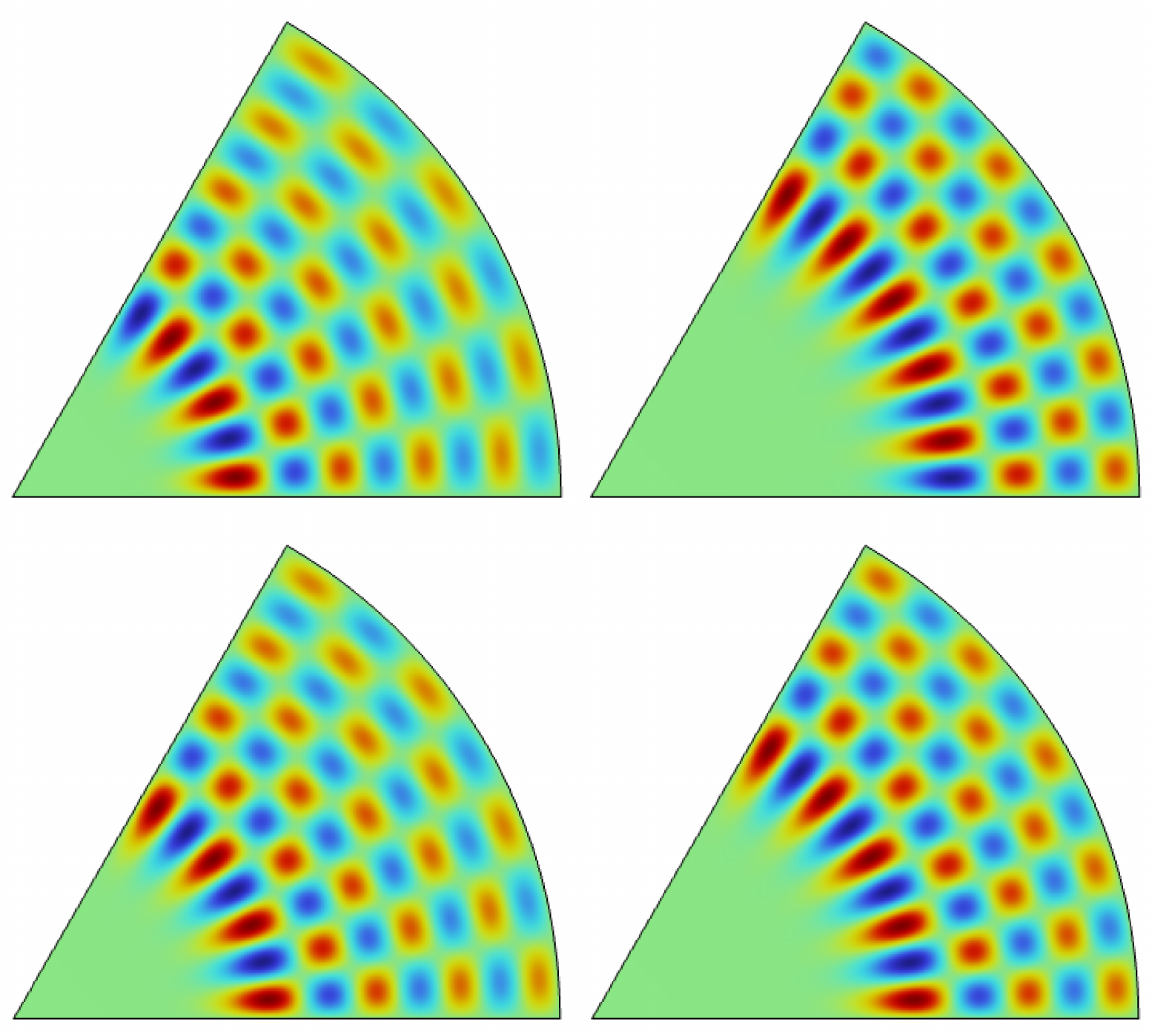}
\includegraphics[width=0.07\linewidth]{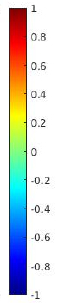}
	\caption{Electric field distributions for the sector-shaped resonator for $\omega < \omega^{crit}$ in the $z=10$~mm plane for, from top left to bottom right, $f=$0.6598~GHz,  0.6625~GHz, 0.6649~GHz, 0.6668~GHz.} 
\label{fig2}
\end{figure}

\begin{figure}[htbp]
\centering
\includegraphics[width=0.5\linewidth]{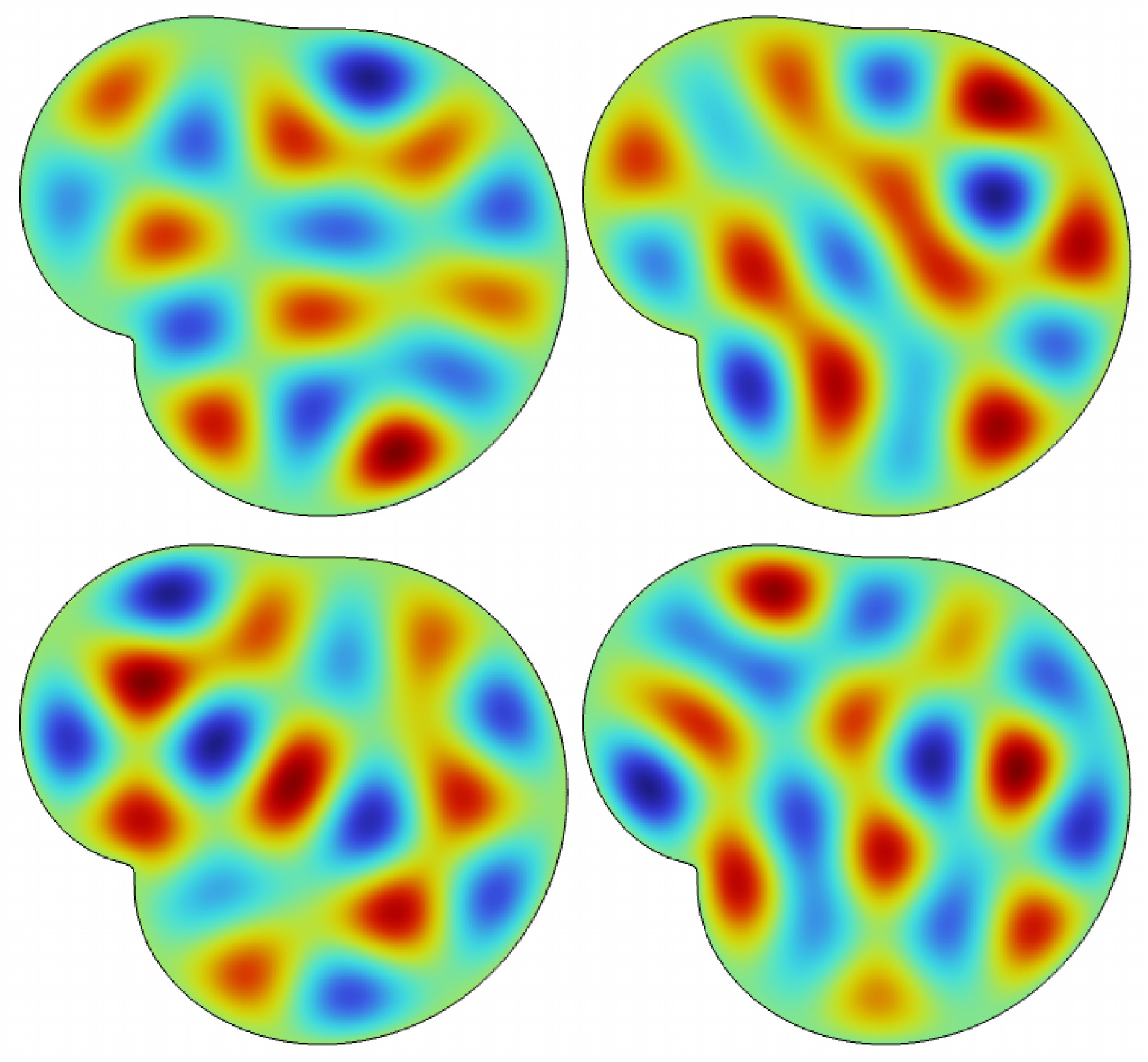}
\includegraphics[width=0.07\linewidth]{Colorbar.pdf}
\caption{Electric field distributions for the Africa-shaped resonator for $\omega < \omega^{crit}$ in the $z=5$~mm plane for, from top left to bottom right, $f=$ 0.9303~GHz, 0.9395~GHz, 0.9540~GHz, 0.9854~GHz.} 
\label{fig3}
\end{figure}

\begin{figure}[htbp]
\centering
\includegraphics[width=0.5\linewidth]{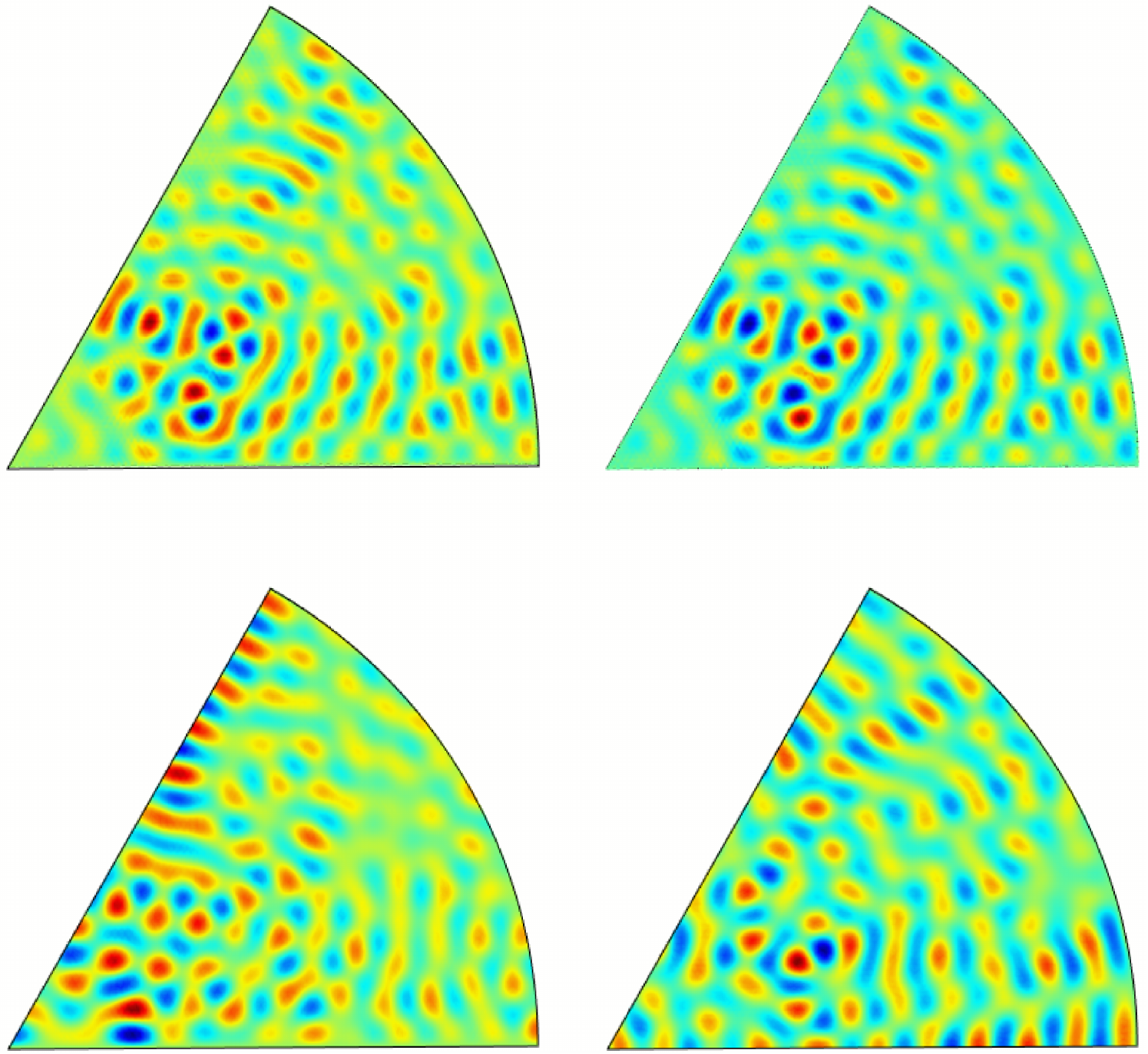}
\includegraphics[width=0.07\linewidth]{Colorbar.pdf}
	\caption{Electric field distribution for the sector-shaped resonator in the $(x,y)$ plane at the eigenfrequency $f=2.0142$~GHz, for which $q=1$. The top left and right figures show $E_z(x,y)$ in the bottom ($z=0$) and top ($z=h$) planes, respectively. The bottom left and right figures show $E_x(x,y)$ and $E_y(x,y)$ in the $z= \frac{h}{2}$ plane.} 
\label{fig4}
\end{figure}

\begin{figure}[htbp]
\centering
\includegraphics[width=0.5\linewidth]{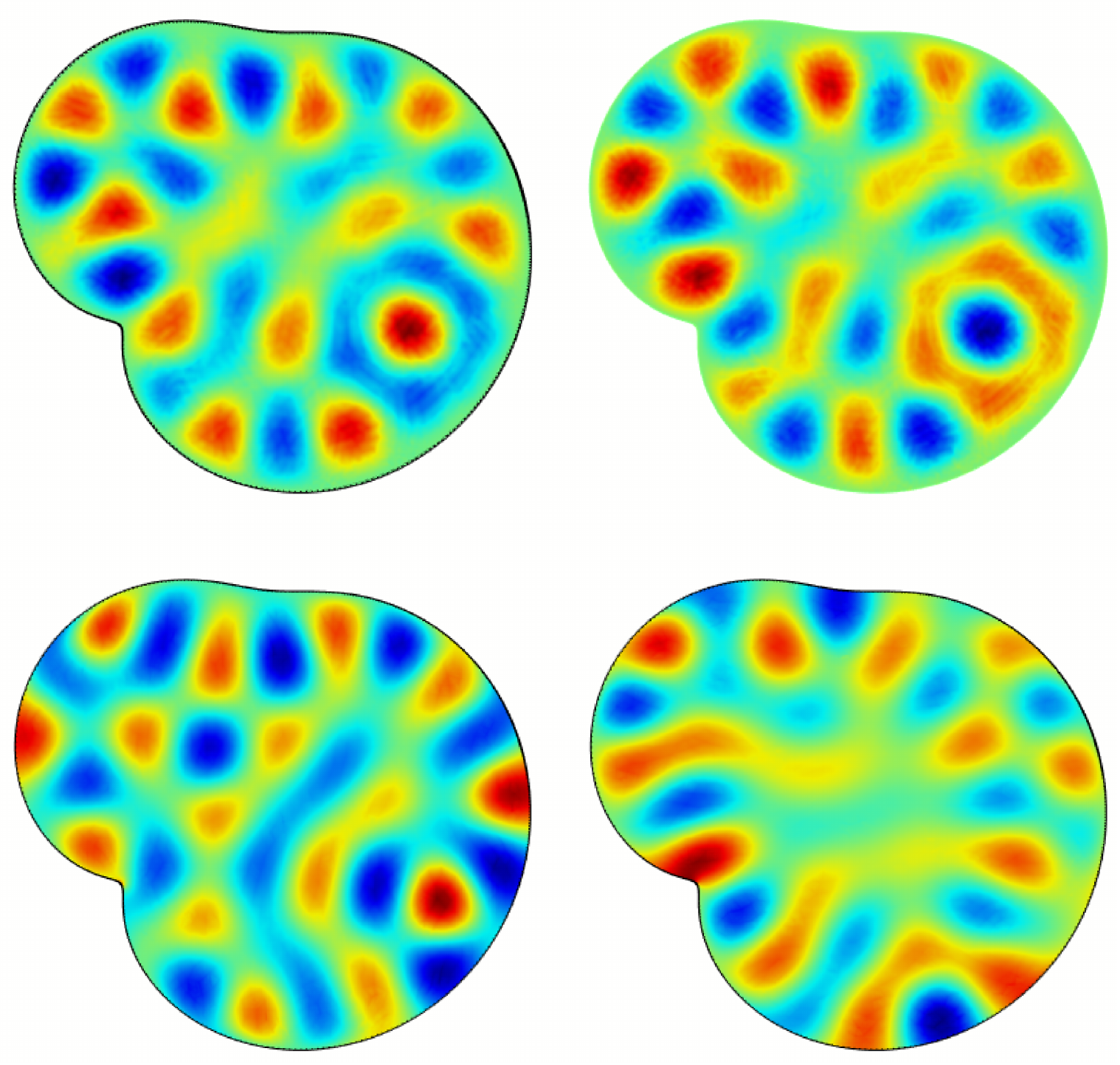}
\includegraphics[width=0.07\linewidth]{Colorbar.pdf}
	\caption{ Electric field distribution for the Africa-shaped resonator in the $(x,y)$ plane at the eigenfrequency $f=2.9652$~GHz, for which $q=1$. The top left and right figures show $E_z(x,y)$ in the bottom ($z=0$) and top ($z=h$) planes, respectively. The bottom left and right figures show $E_x(x,y)$ and $E_y(x,y)$ in the $z= \frac{h}{2}$ plane.}
\label{fig5}
\end{figure}

Below the critical frequency $f^{crit}=\frac{\omega^{crit}}{2\pi}$ corresponding to $k_z=0$, the electric field is perpendicular to the resonator plane $\vec E(x,y)e^{ik_zz}=E(x,y)\vec e_z$. Figures~\ref{fig2} and \ref{fig3} present examples for the electric-field distributions $E(x,y)$ of the sector and Africa resonators for four eigenfrequencies $f=\frac{\omega}{2\pi}$. 
As expected, the electric-field components in the resonator plane, $E_x(\boldsymbol{r})$ and $E_y(\boldsymbol{r})$, are identical to zero, and $E_z(\boldsymbol{r})$ is constant in $z$-direction. This is no longer the case for frequencies beyond the critical frequency, $\omega\gtrsim\omega^{crit}$ where for all components the $z$ dependence is given according to~\refeq{kz} by $e^{-iq\frac{\pi}{h}z}$ with $q\geq 1$ and $\vec E(x,y)$ is governed by the wave equation~\refeq{Helmholtz} together with the boundary conditions~\refeq{BC1} and~\refeq{BC2}. In~\reffig{fig4} and~\reffig{fig5} we show examples for the case $q=1$ for the sector- and Africa-shaped resonators, respectively. In the top and bottom plane the electric field has opposite signs for given values $(x,y)$ as illustrated in the first row of both figures. Furthermore, in the $\left(z=\frac{h}{2}\right)$-plane it vanishes, thus confirming that the $z$ dependence is given by $\sin q\frac{\pi}{h}z$. The second row of both figures shows $\vec E(x,y)$, $E_x(x,y)$ and $E_y(x,y)$ for $z=\frac{h}{2}$. We also confirmed that they fulfill the boundary condition~\refeq{BC1}, that is, vanish in the top and bottom plane. 

\subsection{Spectral Properties}
A central prediction within the field of Quantum Chaos is the Bohigas-Gianonni-Schmit (BGS) conjecture~\cite{Berry1979,Casati1980,Bohigas1984}, which states that for typical quantum systems, whose corresponding classical dynamics is chaotic, the universal fluctuation properties in the eigenvalue spectra coincide with those of random matrices from the Gaussian orthogonal ensemble (GOE) if \T invariance is preserved, and from the Gaussian unitary ensemble (GUE) if it is violated~\cite{LesHouches1989,StoeckmannBuch2000,Mehta2004,Haake2018}. On the other hand, if the classical dynamics is integrable they are well described by uncorrelated random numbers drawn from a Poisson process according to the Berry-Tabor (BT) conjecture~\cite{Berry1977b}. To obtain information on universal fluctuation properties in the eigenfrequency spectra of the ferrite resonators, system-specific properties need to be extracted, that is, the eigenfrequencies have to be unfolded to a uniform average spectral density, respectively, to average spacing unity. Below $\omega^{crit}$ the integrated spectral density is well described by Weyl's formula~\cite{Weyl1912}, as long as the frequency interval is chosen such that the frequency dependence of the dispersion factor in~\refeq{Dispersion} can be neglected. Then, according to Weyl's formula, the smooth part of the integrated spectral density is given by $N^{Weyl}(k)=\frac{\mathcal{A}}{4\pi}k^2+\frac{\mathcal{L}}{4\pi}k+C_0$ with $\mathcal{A}$ and $\mathcal{L}$ denoting the area and perimeter of the resonator shape. Unfolding is achieved by replacing the eigenwavenumbers $k_p$ of~\refeq{Helm2} by the Weyl term $\epsilon_p=N^{Weyl}(k_p)$~\cite{Haake2018}. 

\begin{figure}[htbp]
\centering
\includegraphics[width=0.48\linewidth]{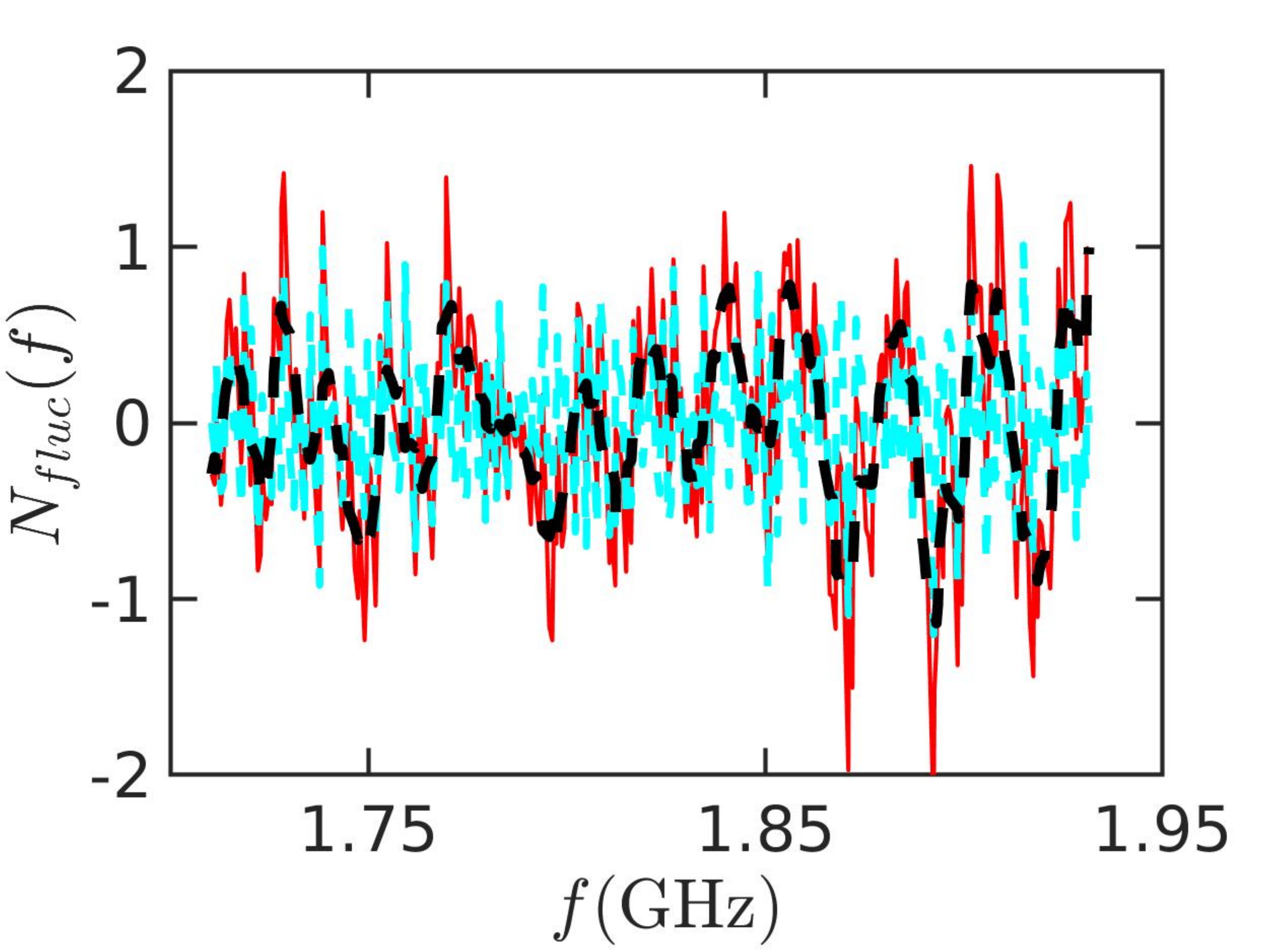}
\includegraphics[width=0.48\linewidth]{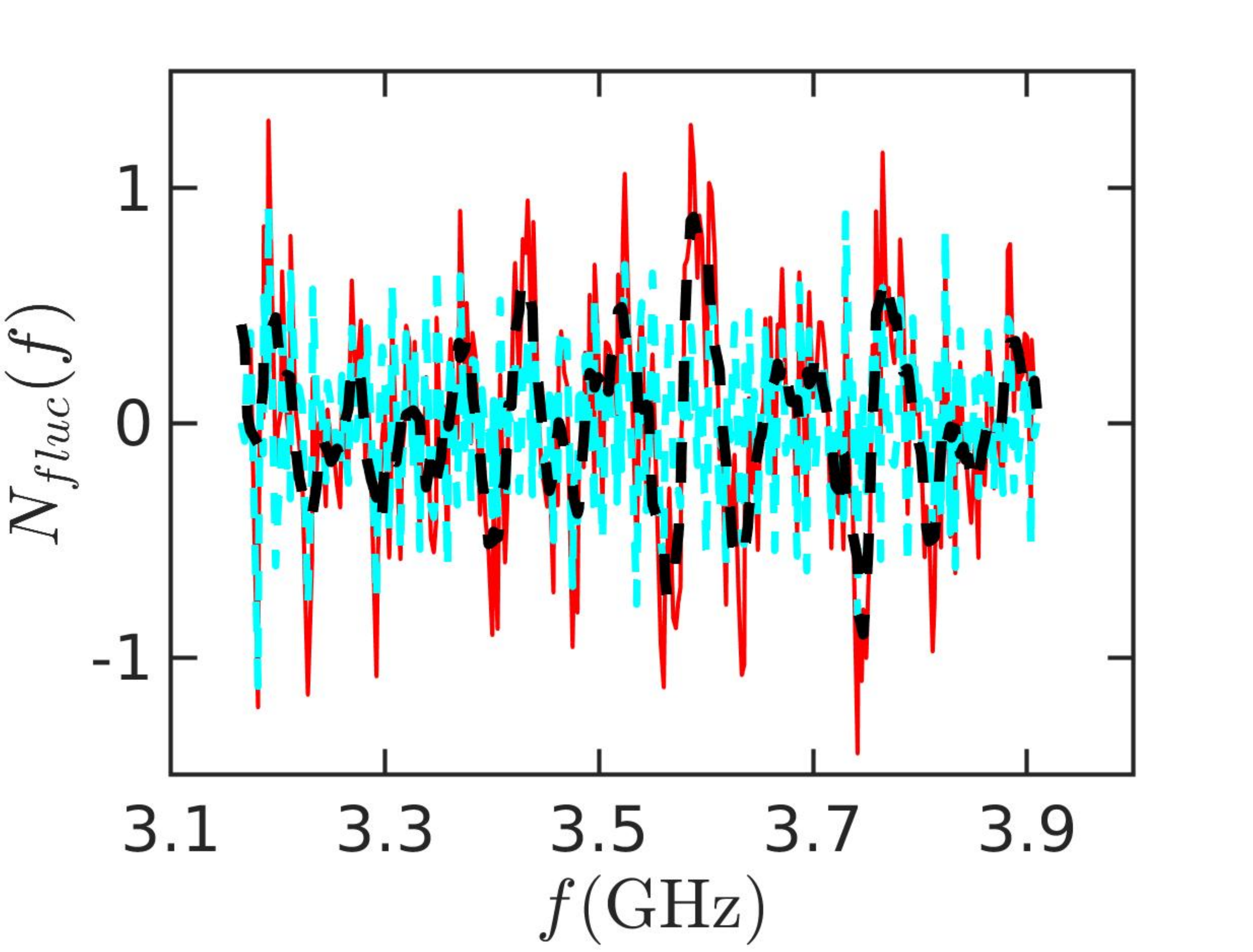}
	\caption{Fluctuating part of the integrated spectral density (red curves) and the slow oscillations (dashed black curve) resulting from bouncing-ball orbits for $\omega\gtrsim\omega^{crit}$ for the sector-shaped (left) and Africa-shaped (right) resonators. The cyan curves show the fluctuations after removal of the contributions from bouncing-ball orbits.}
\label{fig6}
\end{figure}

Above the critical frequency, the Helmholtz equation becomes vectorial. For a three-dimensional metallic cavity with a non-dispersive medium the smooth part of the integrated spectral density is given by a polynomial of third order in $k$~\cite{Balian1970}, where the quadratic term vanishes. For a dispersive medium, like the cavities filled with magnetized ferrite, it still provides a good description of the smooth part of the integrated spectral density, if the frequency range is chosen such that the variation of the dispersion term with frequency is small. In~\reffig{fig6} we show as red solid lines the fluctuating part of the integrated spectral density, $N^{fluc}(k)=N(k)-N^{Weyl}(k)$ for the sector- and Africa-shaped resonators. 
The wave dynamics of a three-dimensional sector-shaped PEC cavity is integrable, whereas the Africa-shaped one comprises non-chaotic bouncing-ball orbits corresponding to microwaves that bounce back and forth between the top and bottom plate~\cite{Alt1996,Dembowski2002,Dietz2008}. These occur in both resonators for $\omega\gtrsim\omega^{crit}$, that is, $k_z\geq\frac{\pi}{h}$. They are non-universal, since they depend on the height of the cavity and lead to deviations from BGS predictions for cavities with otherwise chaotic dynamics, which are similar to those induced by the bouncing-ball orbits in the two- and three-dimensional stadium billiard~\cite{Berry1985,Sieber1993,Dembowski2002,Dietz2008}. The slow oscillations $N^{osc}(k_p)$ in the fluctuating part of the integrated spectral density, depicted as dashed black curves in Fig. \ref{fig6}, originate from these bouncing-ball orbits. We removed them by unfolding the eigenvalues with $\epsilon_p=N^{Weyl}(k_p)+N^{osc}(k_p)$~\cite{Dembowski2002,Dietz2008} which, in addition to the smooth variation, takes into account these oscillations. The resulting  $\tilde N^{fluc}(k_p)=N^{fluc}(k_p)- N^{osc}(k_p)$ is shown as thin dashed turquoise line.

\begin{figure}[htbp]
\centering
\includegraphics[width=0.48\linewidth]{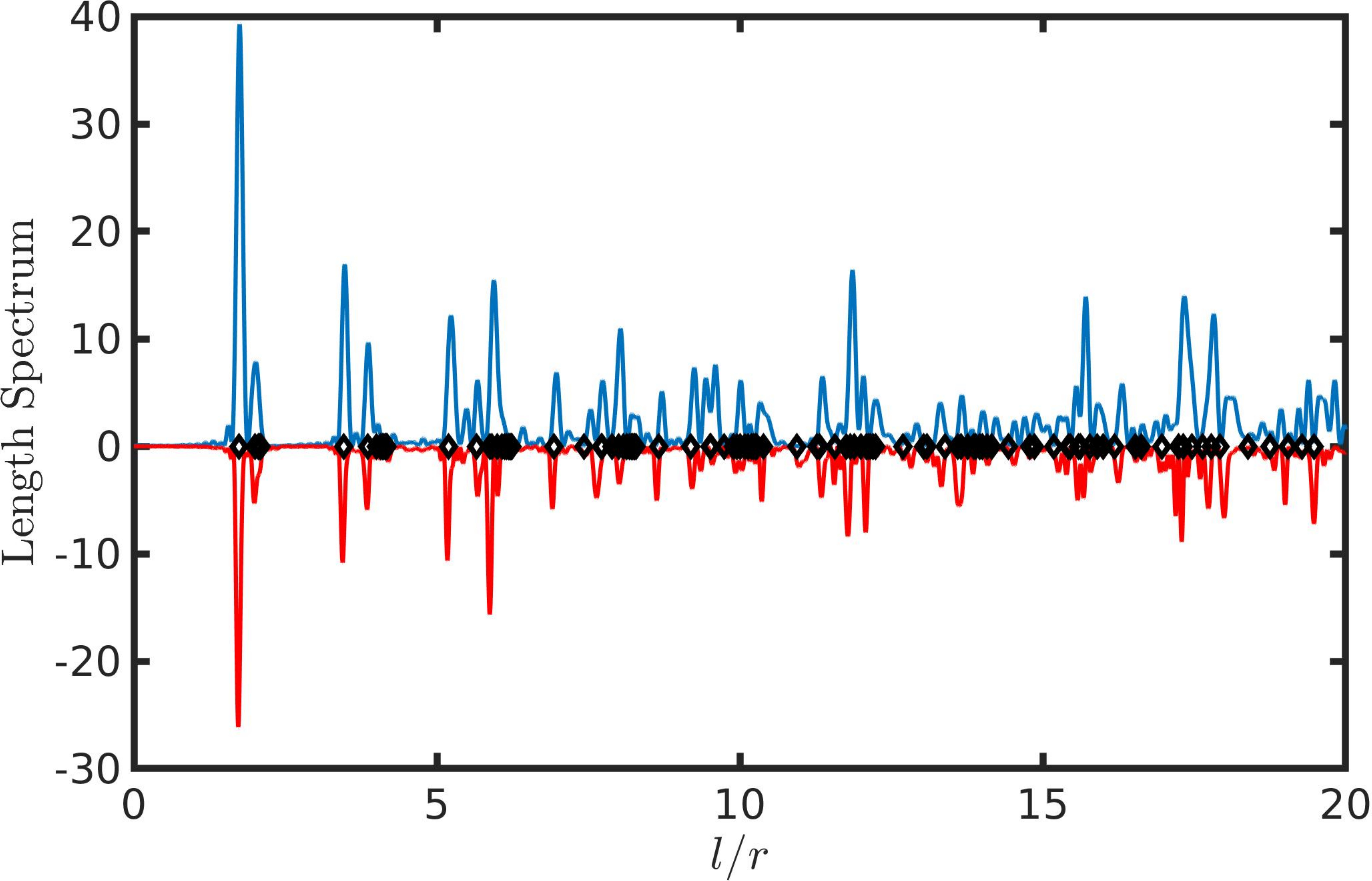}
\includegraphics[width=0.48\linewidth]{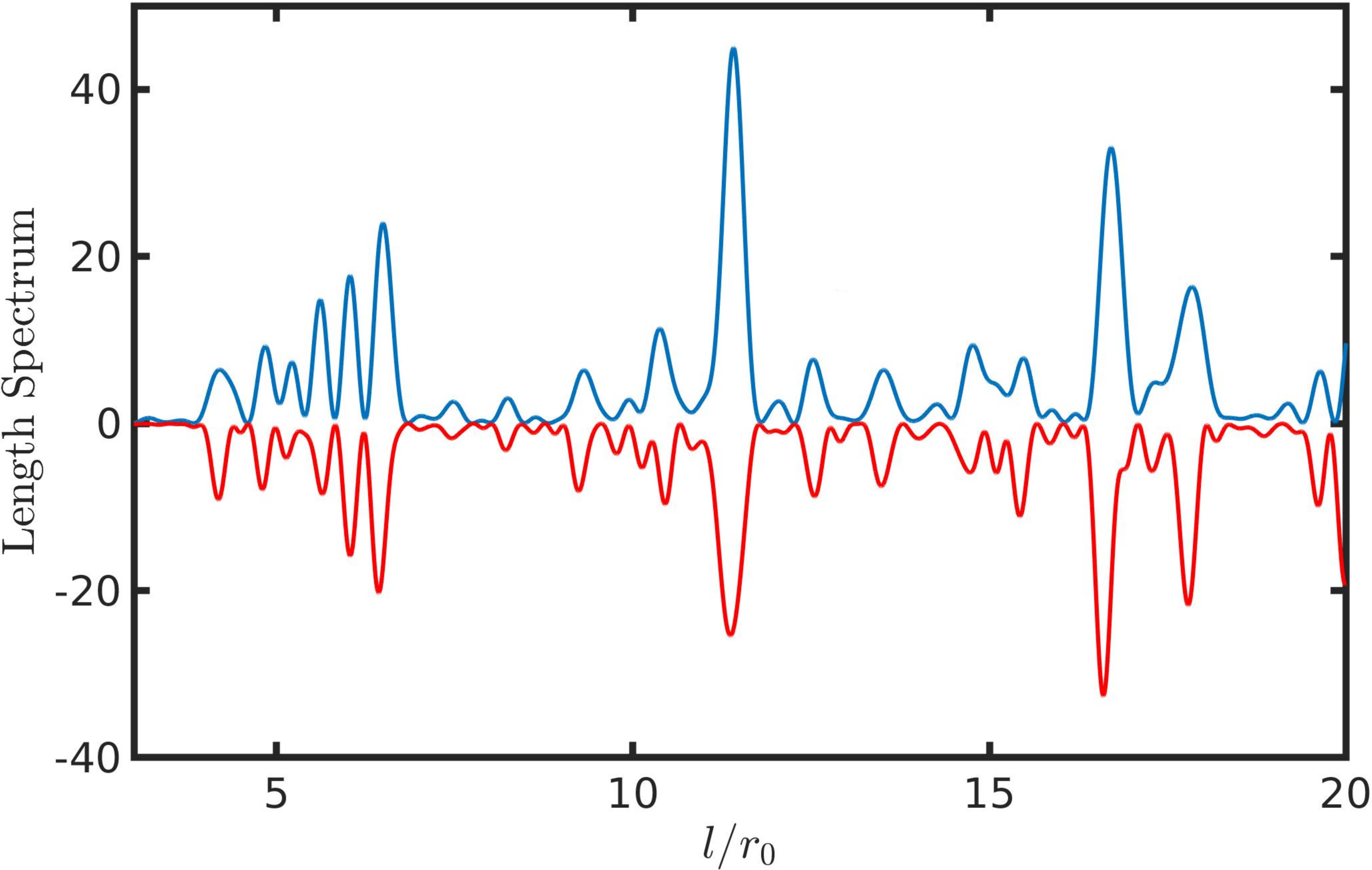}
	\caption{Comparison of the length spectra of the sector-shaped  and Africa-shaped resonators below the critical frequency (turquoise curves) with those obtained from the lowest 300 eigenvalues of the quantum billiard of corresponding shape (red lines). The black diamonds mark the lengths of classical periodic orbits.} 
\label{fig7}
\end{figure}

In~\reffig{fig7} we show length spectra, that is, the modulus of the Fourier transform of the fluctuating part of the spectral density from wavenumber to length. They are named length spectra because they exhibit peaks at the lengths of periodic orbits of the corresponding classical system, as may be deduced from the semiclassical approximation for the fluctuating part of the spectral density~\cite{Berry1976,Gutzwiller1971}. Shown are the length spectra for the sector-shaped (left) and Africa-shaped (right) resonators (turquoise solid lines) below the critical frequency compared to those computed from the eigenvalues of the quantum billiard of corresponding geometry taking into account a similar number of eigenvalues (red solid lines). To match the lengths of the periodic orbits we employed the dispersion relation~\refeq{Dispersion} which provides the relation between the eigenwave numbers of the empty metallic cavity, i.e., the quantum billiard, and ferrite-loaded one. The black diamonds mark the lengths of classical periodic orbits. The agreement between the length spectra is very good, as may be deduced from the fact that below the critical frequency the underlying wave equations are mathematically equivalent. Above the critical frequency this analogy is lost, because of the different structures of the wave equation for an empty metallic cavity~\cite{Jackson1999} and~\refeq{Helmholtz} for a cavity filled with magnetized ferrite and the implicated dispersion relation, which also becomes vectorial.

To study the spectral properties of the ferrite-loaded resonators we analyzed the nearest-neighbor spacing distribution $P(s)$, the integrated nearest-neighbor spacing distribution $I(s)$, the number variance $\Sigma^2(L)$ and the Dyson-Mehta statistic $\Delta_3(L)$, which is a measure for the rigidity of a spectrum~\cite{Bohigas1975,Mehta2004}. Furthermore, we computed distributions of the ratios~\cite{Oganesyan2007,Atas2013} of consecutive spacings between nearest neighbors, $r_j=\frac{\epsilon_{j+1}-\epsilon_{j}}{\epsilon_{j}-\epsilon_{j-1}}$. These are dimensionless implying that unfolding is not required~\cite{Oganesyan2007,Atas2013,Atas2013a}.
\begin{figure}[htbp]
\centering
\includegraphics[width=0.7\linewidth]{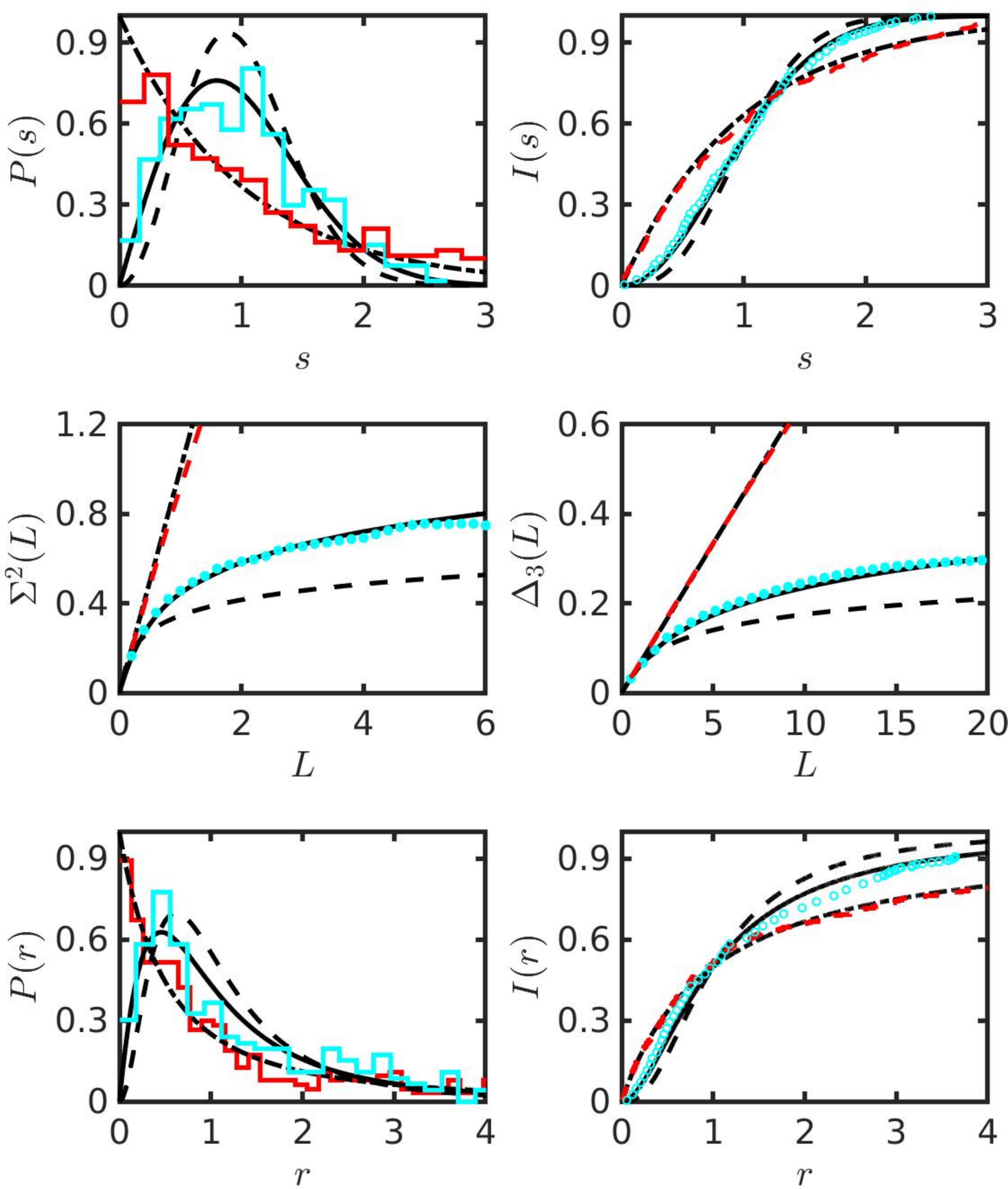}
\caption{Spectral properties of the eigenfrequencies of the sector-shaped resonator below (red histogram and dashed lines) and above (cyan histograms, circles and dots) the critical frequency. They are compared to the spectral properties of Poissonian random numbers (dashed-dotted black lines), GOE (black solid lines), and the GUE (dashed black lines).} 
\label{fig8}
\end{figure} 
\begin{figure}[htbp]
\centering
\includegraphics[width=0.7\linewidth]{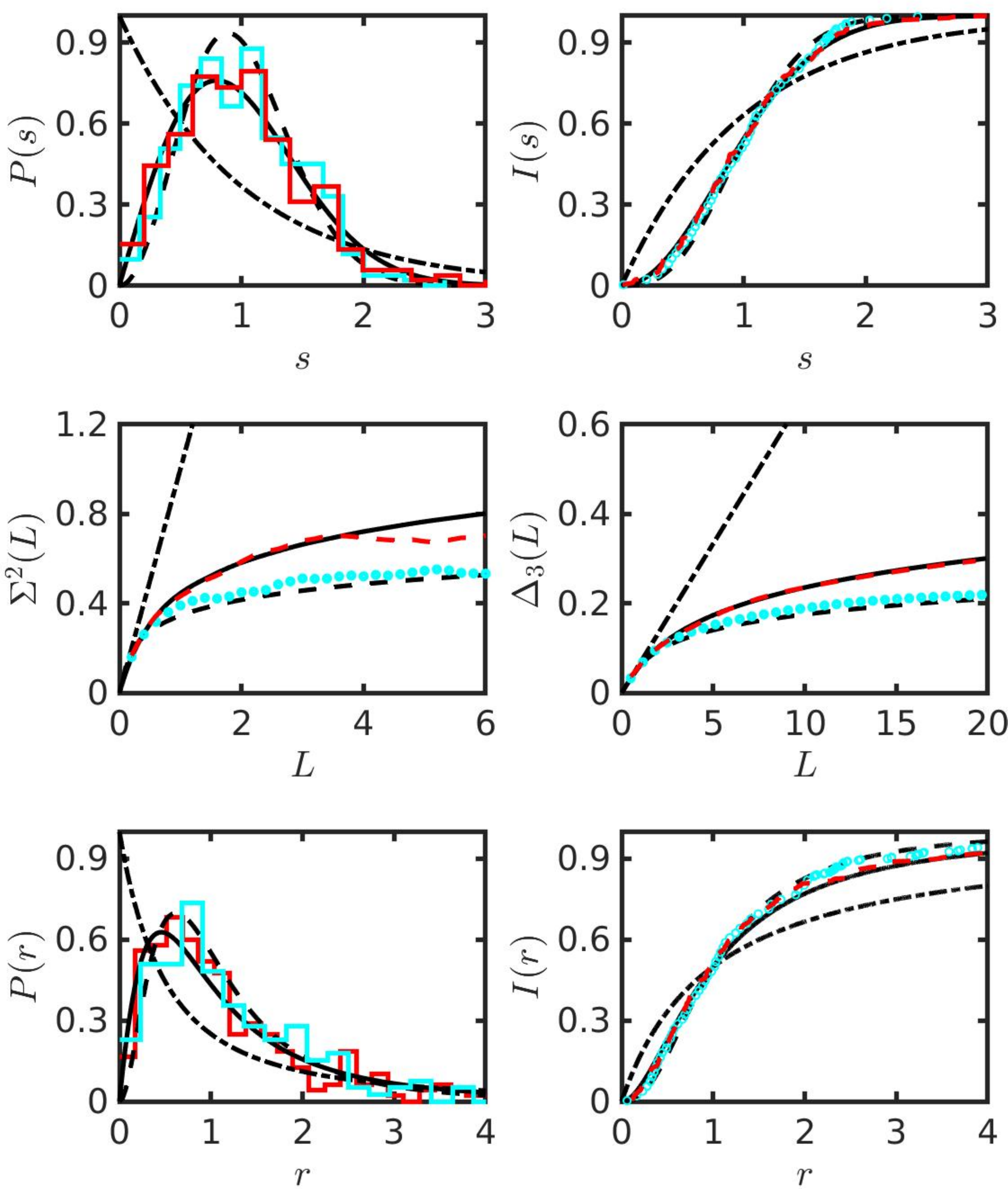}
	\caption{Same as~\reffig{fig8} for the eigenfrequencies of the Africa-shaped resonator.} 
\label{fig9}
\end{figure} 
For the sector-shaped resonator, the spectral properties below (red histograms and dashed lines) and above (cyan histograms, circles and dots) the critical frequency $f^{crit}=1.68$~GHz are shown in~\reffig{fig8}. They agree well with those of Poissonian random numbers, and thus with those of the corresponding quantum billiard below $f^{crit}$ and exhibit GOE statistics in the other case. Here, we used 500 eigenfrequencies in the frequency range $f\in[0.0854,1.5010]$~GHz and 321 eigenfrequencies in the range $f\in[1.7104,1.9314]$~GHz, respectively. For the Africa-shaped resonator, the spectral properties below (red histograms and dashed lines) and above (cyan histograms, circles and dots) $f^{crit}=2.92$~GHz are shown in~\reffig{fig9}, and agree with those of the corresponding quantum billiard, that is with GOE below $f^{crit}$. Above $f^{crit}$ they are well described by the GUE. We used 229 eigenfrequencies in the frequency range $f\in[0.2115,2.6333]$~GHz and 309 eigenfrequencies in the range $f\in[3.1655,3.9094]$~GHz. 
\section{Discussion and Conclusions\label{Concl}}
For both realizations of a PEC resonator loaded with magnetized ferrite, the spectral properties agree with those of the corresponding quantum billiard for $f\lesssim f^{crit}$. Above the critical frequency, the spectral properties of the sector-shaped resonator coincide with those of random matrices from the GOE, implying that there the wave dynamics is chaotic, even though the shape corresponds to that of a three-dimensional billiard with integrable classical dynamics. Above all, the spectral properties of a sector-shaped PEC resonator filled with a homogeneous dielectric exhibit Poissonian statistics~\cite{Lebental2007,Schwefel2005}, that is, their wave dynamics is integrable. Thus we may conclude that the GOE behavior of the sector-shaped resonator and the GUE behavior of the Africa-shaped one have their origin in the magnetization of the ferrite, as may also be concluded from the structure of the wave equation~\refeq{Helmholtz}. It comprises purely complex parts containing derivatives of the entries of $\hat\mu_r$, which are spatially independent in the bulk of the ferrite but experience jumps at the ferrite surface where it is terminated with a PEC. Thereby, the electric-field components of $\vec E(x,y)$ are coupled for non-vanishing static external magnetic field $H_0$, thus leading to the complexity of the dynamics. For $H_0=0$, that is for a dielectric medium, $\hat\mu_r$ equals the identity matrix, so that such a coupling is absent. The spectral properties of the sector-shaped resonator do not exhibit GUE behavior, but are well described by GOE statistics. This is attributed to the mirror symmetry, which implies a generalized \T invariance~\cite{Haake2018}. In the experiments presented in Refs.~\cite{So1995,Wu1998,Schanze2001,Dietz2009,Dietz2010,Dietz2019b} cylindrical ferrites were introduced in a flat, metallic microwave resonator and magnetized with an external magnetic field to induce \Ti-invariance violation. Based on our findings we expect that, when choosing a circular shape of the resonator and inserting the ferrite at the circle center, it acts like a potential which induces wave-dynamical chaos above its critical frequency. XDZ is currently performing such experiments, and preliminary results confirm this assumption. 

\section*{Declarations}
\begin{itemize}
\item {Funding:} {This work was supported by the NSF of China under Grant Nos. 11775100, 12047501, and 11961131009. WZ acknowledges financial support from the China Scholarship Council (No. CSC-202106180044). BD and WZ acknowledge financial support from the Institute for Basic Science in Korea through the project IBS-R024-D1. XDZ thanks the PCS IBS for hospitality and financial support during his visit of the group of Sergej Flach.}
\item {Conflict of Interest:} {The authors declare no conflict of interest.}
\item {Data Avalaibility Statement:} {All data were generated with COMSOL multiphysics under license number 9409425. The parameters to reproduce the data are provided in~\reftab{tab1} and~\refsec{Num}. All data needed to support our findings are included in the figures of this article.}
\item {Authors' Contributions:} {WZ and XZ contributed equally.}
\end{itemize}

\bibliography{References_Ferrite}

\end{document}